\documentclass[traditabstract]{aa}

\usepackage{graphicx}
\usepackage{amssymb}
\usepackage{amsmath}
\usepackage{url}
\usepackage{natbib}
\bibpunct{(}{)}{;}{a}{}{,}

\newcommand{\onefig}[3]{%
  \begin{figure}%
    \centerline{\resizebox{\hsize}{!}{\includegraphics*{#1}}}%
    \caption{#3}\label{#2}%
  \end{figure}%
}
\newcommand{\twofig}[4]{%
  \begin{figure*}%
    \centerline{\resizebox{\hsize}{!}{\includegraphics*{#1} \,%
        \includegraphics*{#2}}}%
    \caption{#4}\label{#3}%
  \end{figure*}%
}

\newcommand{\fourfig}[6]{%
  \begin{figure*}%
    \centerline{\resizebox{\hsize}{!}{\includegraphics*{#1} \,%
        \includegraphics*{#2}}} \,%
    \centerline{\resizebox{\hsize}{!}{\includegraphics*{#3} \,%
        \includegraphics*{#4}}}%
    \caption{#6}\label{#5}%
  \end{figure*}%
}
\newcommand{\sect}[1]{Section~\ref{#1}}
\newcommand{\fig}[1]{Figure~\ref{#1}}
\newcommand{\eq}[1]{Equation~(\ref{#1})}
\newcommand{\tab}[1]{Table~\ref{#1}}
\newcommand\lmax{\ell_\mathrm{max}}
\newcommand\npix{n_{\mathrm{pix}}}
\newcommand\nside{n_{\mathrm{side}}}
\newcommand\alm{a_{\ell m}}
\newcommand{\order}[1]{$\mathcal{O}({#1})$}
\newcommand\nv{\hat{n}}
\newcommand\libpsht{\texttt{libpsht}}
\newcommand\beq{\begin{equation}}
\newcommand\eeq{\end{equation}}

\begin{document}

\title{ARKCoS: Artifact-suppressed accelerated radial kernel
  convolution on the sphere}

\titlerunning{ARKCoS: Accelerated radial kernel convolution on the
  sphere}

\author{Franz Elsner\inst{1},
  Benjamin D. Wandelt\inst{1,2}}

\offprints{elsner@iap.fr}

\institute{Institut d'Astrophysique de Paris, UMR 7095, CNRS -
  Universit\'e Pierre et Marie Curie (Univ Paris 06), 98 bis blvd
  Arago, 75014 Paris, France
  \and
  Departments of Physics and Astronomy, University of Illinois at
  Urbana-Champaign, Urbana, IL 61801, USA}

\date{Received \dots / Accepted \dots}

\abstract {We describe a hybrid Fourier/direct space convolution
  algorithm for compact radial (azimuthally symmetric) kernels on the
  sphere. For high resolution maps covering a large fraction of the
  sky, our implementation takes advantage of the inexpensive massive
  parallelism afforded by consumer graphics processing units
  (GPUs). Its applications include modeling of instrumental beam
  shapes in terms of compact kernels, computation of fine-scale
  wavelet transformations, and optimal filtering for the detection of
  point sources. Our algorithm works for any pixelization where pixels
  are grouped into isolatitude rings. Even for kernels that are not
  bandwidth-limited, ringing features are completely absent on an ECP
  grid. We demonstrate that they can be highly suppressed on the
  popular HEALPix pixelization, for which we develop a freely
  available implementation of the algorithm. As an example
  application, we show that running on a high-end consumer graphics
  card our method speeds up beam convolution for simulations of a
  characteristic Planck high frequency instrument channel by two
  orders of magnitude compared to the commonly used HEALPix
  implementation on one CPU core, while typically maintaining a
  fractional RMS accuracy of about 1 part in $10^{5}$.}


\maketitle

\section{Motivation and goals}

Convolving with radial (i.e.\ azimuthally symmetric) kernels is a key
step in some of the most frequently used algorithms during the
simulation and analysis of cosmological data sets represented on the
celestial sphere, such as maps of the cosmic microwave background
(CMB).

All current and future CMB experiments have many ($100$--$10^{4}$)
detectors (e.g., the Atacama Cosmology Telescope,
\citealt{2003NewAR..47..939K}, the South Pole Telescope,
\citealt{2004SPIE.5498...11R}, the proposed CMBPol mission,
\citealt{baumann:10}, or the Planck satellite,
\citealt{2011arXiv1101.2022P}). Simulating the signal in these data
sets requires very many beam smoothing operations since each detector
map will contain the same CMB map smoothed with a separate beam
shape. The same is true for map-making methods that compute the
optimal combination of a large number of detectors in an iterative
process and therefore also require a huge number of beam smoothing
operations \citep[e.g.,][]{1997ApJ...480L..87T, 2001A&A...372..346N,
  2002PhRvD..65b2003S}.

Several CMB analysis techniques, such as a wavelet analysis
\citep[e.g.,][]{1999MNRAS.309..125H, 2002MNRAS.336...22M,
  2004ApJ...609...22V}, and the filtering to detect point sources
\citep[e.g.,][]{1998ApJ...500L..83T, 2000MNRAS.315..757C,
  2006MNRAS.369.1603G} require smoothing of high resolution maps with
symmetric kernels that have (or are well-approximated as having)
compact support on the sphere. In a wavelet analysis, the
computational time for the wavelet transform is dominated by the
computation of the fine-scale wavelet coefficients. By construction,
the fine-scale wavelets are compact in pixel space.

Current practice in CMB data analysis is the near-exclusive use of the
fast spherical harmonic transform (FSHT) for convolution with radial
kernels \citep{1997ApJ...488L..63M}. Mature and highly efficient
implementations of this algorithm are publicly available in several
packages, such as, e.g.,
HEALPix\footnote{\url{http://healpix.jpl.nasa.gov}}
\citep{2005ApJ...622..759G},
GLESP\footnote{\url{http://www.glesp.nbi.dk}}
\citep{2005IJMPD..14..275D},
ccSHT\footnote{\url{http://crd.lbl.gov/~cmc/ccSHTlib/doc}}, or
\libpsht\footnote{\url{http://sourceforge.net/projects/libpsht}}
\citep{2011A&A...526A.108R}.

In the vast majority of cases that a spherical transform is calculated
during a CMB data analysis, it is to compute a convolution with a
radial kernel. Examples are: (1) simulating CMB maps, in which case
the radial kernel is the ``square root'' of the CMB power spectrum;
(2) simulating observed detector maps with a symmetric beam profile;
(3) filtering to extract point sources, or hot and cold spots on
certain scales; and (4) all forms of symmetric wavelet analysis.
 
While generally correct, this approach is not optimal when convolving
high resolution maps with sufficiently compact kernels. We show that
significant speed-up is possible with an algorithm that makes nearly
optimal use of massively parallel fast Fourier transforms (FFT) on
consumer graphical processing units using a hybrid direct-space and
Fourier space approach.

The plan of this paper is as follows. We begin with a mathematical
definition of the problem and the quantities involved
(\sect{sec:intro}). We then briefly review the convolution approaches
using a direct sum and the fast spherical harmonic transform in
\sect{sec:methods}, while discussing a GPU implementation of the FSHT
\citep{2010arXiv1010.1260H}. We introduce our algorithm and its
implementation on a consumer GPU in \sect{sec:hybrid}, which also
contains benchmark results and tests of the numerical accuracy of the
algorithm. Finally, we summarize our findings in
\sect{sec:conclusions}.

The benchmarks of the algorithms reported in this paper were performed
on an Intel Core2 Quad CPU with 2.8~GHz and 8 GB of random access
memory (RAM). The system cost (other than the GPU) was about
US\$~1000. As a reference, we use the popular HEALPix Fortran package
version 2.15 and the highly optimized \libpsht\ C++ FSHT library. We
note that starting with the latest release, version 2.20, the
\libpsht\ routines will also be called by default in the HEALPix
package. Our GPU code was timed on a NVIDIA GeForce GTX 480 that we
bought for US\$~500.

\section{Definitions and notation}
\label{sec:intro}

It is useful to state the "platonic ideal" of what is to be
accomplished. Given a rough map $r$, we would like to calculate the
smooth map $s$
\beq
s({\nv_1})=\int_{S^2} K({\nv_1},{\nv_2}) r({\nv_2}) d^2{\nv_2},
\label{eq:def}
\eeq
where ${\nv}$ denotes a unit vector on the sphere. For a symmetric
(or \textit{radial}) kernel, $K({\nv}_1,{\nv}_2) =K({\nv}_1.{\nv}_2)$.
We introduce the short-hand notation $p.q \equiv
{\nv_p}.{\nv_q}=\cos\left(\sphericalangle({\nv}_p,{\nv}_q)\right)$,
so $K({\nv_1},{\nv_2})= K(1.2)$.
 
A band-limited function on the sphere can be defined in spherical
harmonic space by specifying a set of spherical harmonic coefficients
$\alm$ for all $\ell$ from zero up to band-limit $\lmax$. Unless
otherwise stated sums are over all non-zero terms. With the Legendre
transform convention
\beq 
K_\ell=2\pi \int_{-1}^{1} K(z) P_\ell(z) dz \, ,
\eeq
 the kernel can be expanded in terms of Legendre polynomials as
\beq
K(p.q)=\sum_{\ell} \frac{2\ell+1}{4\pi}K_\ell P_\ell(p.q) \, .
\eeq 
We assume that the kernel has the same band-limit as the input map.
Recalling the addition theorem for spherical harmonics $Y_{\ell m
  p}\equiv Y_{\ell m} ({\nv_p})$
\beq
\sum_m Y_{\ell m p} Y^\ast_{\ell m q}=\frac{2\ell+1}{4\pi}P_\ell(p.q) \, ,
\eeq
we obtain
\beq
s_{\ell m} = K_\ell\, r_{\ell m} \, .
\label{eq:lspacedef}
\eeq
This equation is exact if the $r_{\ell m}$ are known. In many cases of
interest, however, the map will be available in a sampled or pixelized
representation with a number of pixels $\npix$. In this case,
estimating the $r_{\ell m}$ from the sampled representation may
introduce a quadrature error. We keep this in mind when discussing the
convolution accuracy in the following.

\section{Methods}
\label{sec:methods}

\subsection{Direct sum}

The direct sum follows from the straightforward discretization of
\eq{eq:def}
\beq
\label{eq:direct}
s_p=\sum_q K(p.q) r_q \, .
\eeq
It is easy to check that this approach will yield the same output map
as \eq{eq:lspacedef} if the $r_{\ell m}$ are calculated by direct sum
over the same equal-area pixelization and both map and radial kernel
are band-limited functions. In general, this method scales as
\order{\npix^2\sim\lmax^4}, both in terms of memory accesses and in
terms of floating point operations (FLOP). The prefactor can be made
small by caching $K(z)$, e.g.\ by interpolating it in \order{1}
operations from \order{\lmax} precomputed values. This also reduces
the number of accesses to non-cached memory to \order{\lmax^2}.

If the kernel is compact such that $K(p.q) = 0 \ \forall p.q <
z_{\mathrm K}$, i.e.\ for an angle between $p$ and $q$ larger than a
threshold $\theta_{\mathrm K}$, the operation count reduces by a
factor $z_{\mathrm K}/2$. For sufficiently compact kernels this method
would therefore win over other methods with smaller asymptotic
time complexity but larger prefactors.

The direct sum has a great degree of parallelism, at least at the
level of \order{\npix\sim\lmax^2} threads, since each output pixel is
the result of a dot product of one row of $K$ with $r$. Theoretically,
even more parallelization can be achieved by parallelizing the dot
product, though care must be taken to avoid race conditions when
accumulating the smoothed map in parallel. In practice, care must be
taken to keep the number of non-cached memory accesses low since the
computation would otherwise be limited by memory bandwidth. Since the
number of memory accesses is of the same order as the number of
calculations, the potential for GPU implementation is limited. Direct
pixel space convolution will therefore be superior only for kernels
that are too small (of widths narrower than a very small number of
pixels) to be of broad practical interest.

\subsection{Fast spherical harmonic transform}

Pixelizations consisting of uniformly sampled isolatitude rings allow
for a fast spherical harmonic transform (FSHT), with overall scaling
\order {\lmax^{3} } to take advantage of \eq{eq:lspacedef}.

In detail, FFTs on the \order{n_{\theta} \sim \lmax} isolatitude rings
(each containing \order{n_\phi \sim \lmax} pixels) are done in
\order{\lmax^2 \log \lmax} operations. The resulting Fourier
components $b_m(\theta)$ can be transformed into spherical harmonic
coefficients $r_{\ell m}$ by applying an associated Legendre transform
taking \order{\lmax^2} operations per ring, for a total of
\order{n_{\theta} \lmax^2 \sim \lmax^3} operations. One obtains the
smoothed map by multiplying the $r_{\ell m}$ with the kernel
coefficients $K_{l}$ and applying the inverse spherical harmonic
transform that inverts the above steps in reverse order again taking
\order{\lmax^{3}} operations.

The Legendre transforms therefore dominate the scaling since applying
\eq{eq:lspacedef} takes only \order{\lmax^2} time. Furthermore, at
high $\ell$, the recursions necessary to compute the associated
Legendre functions become increasingly less accurate and need to be
done in double precision with frequent rescaling to avoid floating
point underflows \citep{2005ApJ...622..759G}. Implementing the FSHT
algorithm on consumer GPUs with reduced double precision performance
is therefore non-trivial. The inverse spherical harmonic transform was
implemented and benchmarked citing \order{10} speed gains with respect
to the
S2HAT\footnote{\url{http://www.apc.univ-paris7.fr/~radek/s2hat.html}}
CPU implementation \citep{2010arXiv1010.1260H}.

The FSHT algorithm is popular since it reduces the computational
scaling by a factor $\lmax$ compared to the direct sum and yet yields
the same result (for infinite precision arithmetic). Approximating the
continuous spherical harmonic transform by a finite, discrete sum over
pixels introduces the same error as approximating the kernel integral
by such a discrete sum. If a quadrature rule is applied to improve the
accuracy of the $r_{\ell m}$, the same quadrature weights can be used
in \eq{eq:direct} to reach the identically improved results.

Utilizing the popular HEALPix library on a modern CPU, the serial time
for a pair of forward and inverse transforms at Planck resolution
($\nside = 2048, \ \lmax = 4096$) is $t \approx 460$~s. Since using
FSHTs is by far the most common algorithm of choice for symmetric
kernel convolutions, this is the reference time for comparisons with
other algorithmic approaches.

Methods implementing divide-and-conquer schemes for fast transforms on
non-Abelian groups have a smaller asymptotic scaling of CPU time with
problem size than the FSHT \citep{184069.184073D,
  2007JCoPh.226.2359W}. However, these more sophisticated methods
require large amounts of RAM to store precomputed information that
renders them impractical for problem sizes of interest for CMB maps,
with tens of millions of pixels e.g.\ from Planck. For smaller problem
sizes, the comparatively large complexity of the algorithm causes
actual implementations to be slower than algorithmically simpler
approaches.

\section{Hybrid method}
\label{sec:hybrid}

We now outline a straightforward hybrid method that combines aspects
of the direct summation and spherical harmonic transformation
approach. It is based on the simple idea of convolving along
isolatitude rings via computationally inexpensive FFTs by means of the
convolution theorem, and integrating in the longitudinal direction in
pixel space.

This hybrid method is redundant in a way that the product of the
kernel image and the input map must be evaluated once on every ring
prior to the summation. The computational costs amount to \order{\lmax
  \log \lmax} operations for each FFT on a ring, which must be
repeated \order{n_{\theta} \sim \lmax} times for each of the
\order{n_{\theta} \sim \lmax} rings. In total, the algorithm requires
\order{\lmax^3 \log \lmax} operations, formally inferior to the
conventional FSHT approach. Unlike that method, however, if the
convolution kernel has finite support on only $n_{\mathrm{support}} <
n_{\theta}$ rings, the computational complexity decreases linearly,
\order{n_{\mathrm{support}} \, \lmax^2 \log \lmax}. It is dominated by
FFTs for which highly optimized implementations with a small prefactor
exist. Furthermore, the algorithm intrinsically offers an extreme
amount of data parallelism, making it in particular suitable for an
implementation on GPUs with hundreds of cores.

In practice, the algorithm field-of-application is limited to compact
kernels. If a kernel is formally non-zero across the entire sphere,
but vanishes sufficiently fast beyond a given angular distance
$\alpha_{\mathrm{cut}}$, it can be truncated at that radius without
introducing significant errors. For the convolution of an isotropic
map with power spectrum $\mathcal{C}_{\ell}$, the mean quadratic error
introduced by this approximation can be estimated to be
\beq
\sigma^2 = \sum_{\ell = 0}^{\lmax} \frac{(2\ell + 1)}{4\pi} \, \Delta
K_{\ell}^2 \, \mathcal{C}_{\ell} \, ,
\eeq
where $\Delta K_{\ell}$ is the Legendre expansion of the difference of
the exact and the truncated kernel.

\subsection{Overview of the algorithm}

We now describe the GPU implementation of the convolution algorithm
for an input map in HEALPix format in greater detail. We visualize the
individual steps of the algorithm in \fig{fig:method}.

HEALPix maps with resolution parameter $\nside$ are divided into three
regions, the north polar cap, the equatorial region, and the south
polar cap. Each of the two caps consist of $n_{\mathrm{caps}} = \nside
- 1$ rings, where the $n$th ring (counted from the pole) contains $4
\, n$ pixels. The equatorial region comprises $n_{\mathrm{equ}} = 2 \,
\nside + 1$ rings with a fixed number of $4 \, \nside$ pixels per
ring.

We perform a real-to-complex Fourier transform of length
$n_{\mathrm{FFT}}$ on each ring, where $n_{\mathrm{FFT}} = 4 \,
\nside$ in the equatorial region but only $n_{\mathrm{FFT}} = 4 \, n$
in the polar caps. We then zero-pad the Fourier coefficients around
the poles to generate a rectangular array of $4 \, \nside - 1$ sets of
$2 \, \nside + 1$ Fourier coefficients each. The rectangular shape of
this array allows us to use the batch FFT mode of our FFT library for
the Fourier convolutions that gives us a significant time saving
since FFTs dominate our computational time budget. As every ring in
the polar and every other ring in the equatorial region of a HEALPix
map is shifted by $\phi_{\mathrm{0}} = \pi/(4 \, n)$ and
$\phi_{\mathrm{0}} = \pi/(4 \, \nside)$, respectively
\citep{2005ApJ...622..759G}, we compensate for this distortion by
phase-shifting the $m$th Fourier coefficient
\beq
\label{equ:fftrot}
b_{\mathrm{m}}'(\theta) = b_{\mathrm{m}}(\theta)
\mathrm{e}^{i \, m \phi_{\mathrm{0}}} \, .
\eeq

After preparing the input map in this way, we can start the
convolution process, which loops over all rings in the output map. For
each ring at latitude $\theta_{\mathrm{0}}$, we generate a kernel grid
of size $n_{\mathrm{support}} \times 4 \, \nside$ pixels and place the
kernel at its center at $(\theta, \phi) = (\theta_{\mathrm{0}},
0)$. To finally interpolate the kernel on this grid, we first
calculate the angular distance $\alpha$ between a pixel located at
$(\theta, \phi)$ and the kernel. A Taylor expansion to second order in
$\alpha$ allows us to rewrite the equation in a simplified way
\begin{align}
  \label{eq:kernel_angle}
  \alpha^2 &\approx 2 \left[ \sin^2(1/2(\theta + \theta_{\mathrm{0}}))
  + \sin^2(1/2(\theta - \theta_{\mathrm{0}})) \right] \nonumber \\
  & \quad - 2 \left[ \sin^2(1/2(\theta + \theta_{\mathrm{0}}))
  - \sin^2(1/2(\theta - \theta_{\mathrm{0}})) \right] \cos(\phi) \, .
\end{align}
Using this parametrization of the angular distance has the advantage
of making a numerically expensive arccosine operation
redundant. Though \eq{eq:kernel_angle} is formally only applicable in
the small angle regime, we can exactly compensate for the error by
systematically biasing the kernel evaluation. To this end, we first
define $\epsilon(\alpha) \equiv \cos(\alpha) - (1 - 1/2 \, \alpha^2)$
as the error introduced by the approximation. Instead of precomputing
a table of the form $\left[ \alpha^2, K(\alpha) \right]$, we then
store a datastructure containing the values $\left[ \alpha^2, K(\alpha
  + \epsilon(\alpha)) \right]$. To evaluate the kernel, we interpolate
linearly from the table.

After interpolating the kernel on the grid, we perform an efficient
radix-2 batch FFT. The kernel Fourier coefficients are then multiplied
by those of the input map. As a next step, for each ring pixel, we
calculate the summation over $n_{\mathrm{support}}$ elements in the
$\theta$-direction. Finally, we store the output-ring Fourier
coefficients and continue to the next ring.

To transform back to a HEALPix map, we reverse phase-shift the rings
and perform a complex-to-real backward FFT on all output rings. In
polar rings, we truncate the Fourier expansion to the Nyquist
frequency of each individual ring. An alternative procedure would have
been to explicitly alias the super-Nyquist modes to sub-Nyquist modes
for each ring. Our tests show that this produces negligible
differences for maps with a band-limit $\lmax \le 2 \, \nside$.

Our approach avoids the recursions for the associated Legendre
functions. Therefore, the entire algorithm could in principle be
computed in single precision, which renders it particularly suitable
for inexpensive consumer GPUs that have limited double precision
capability. However, for very narrow or highly varying kernel in
particular, calculating the angular distance between a given pixel and
the kernel center, \eq{eq:kernel_angle}, becomes imprecise. We
therefore compute this quantity using double precision arithmetic for
the entire calculation. Further performance improvements at the 10~\%
level can be realized by partially relaxing this requirement for
sufficiently smooth kernels or less stringent accuracy goals.

\fourfig{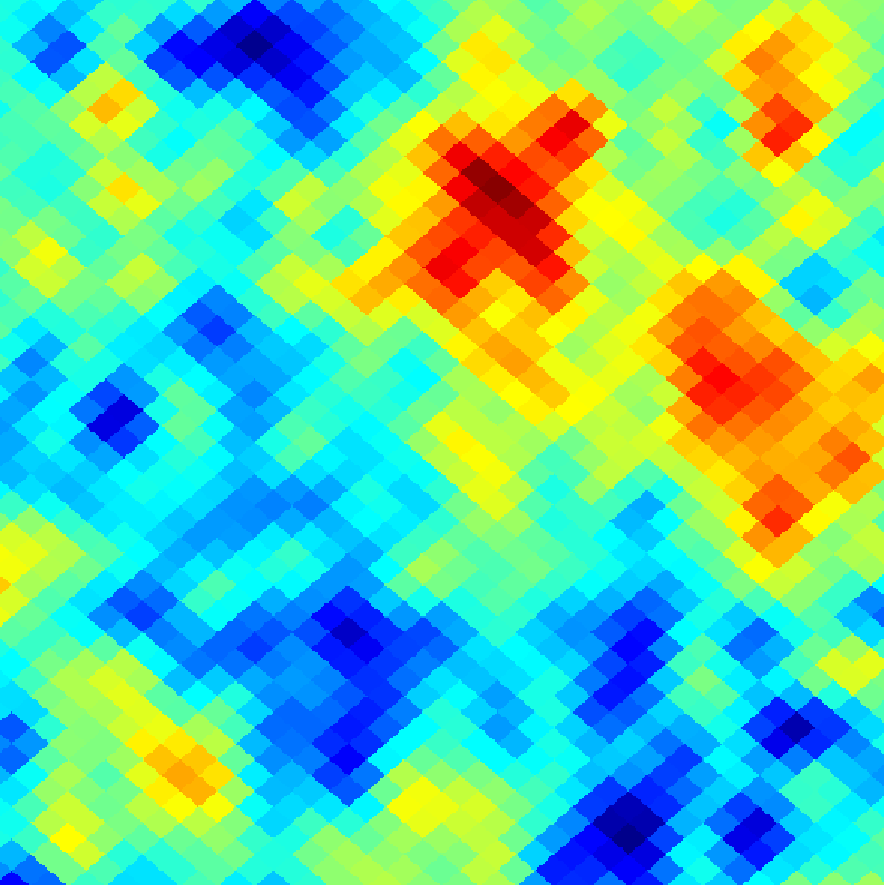}{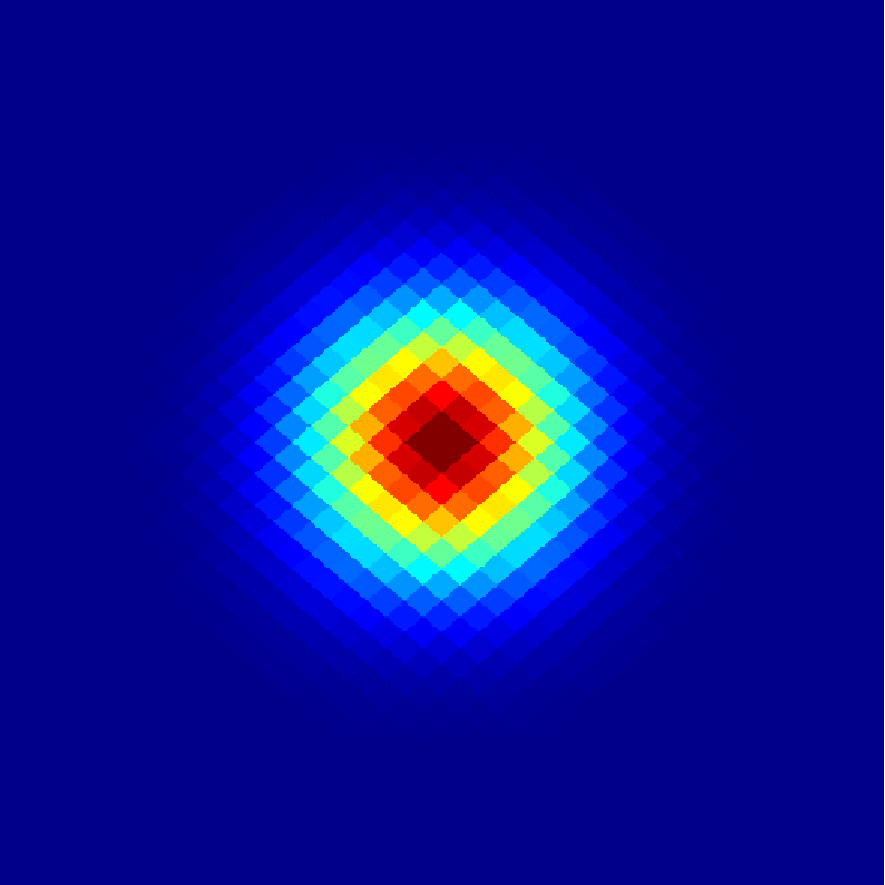}{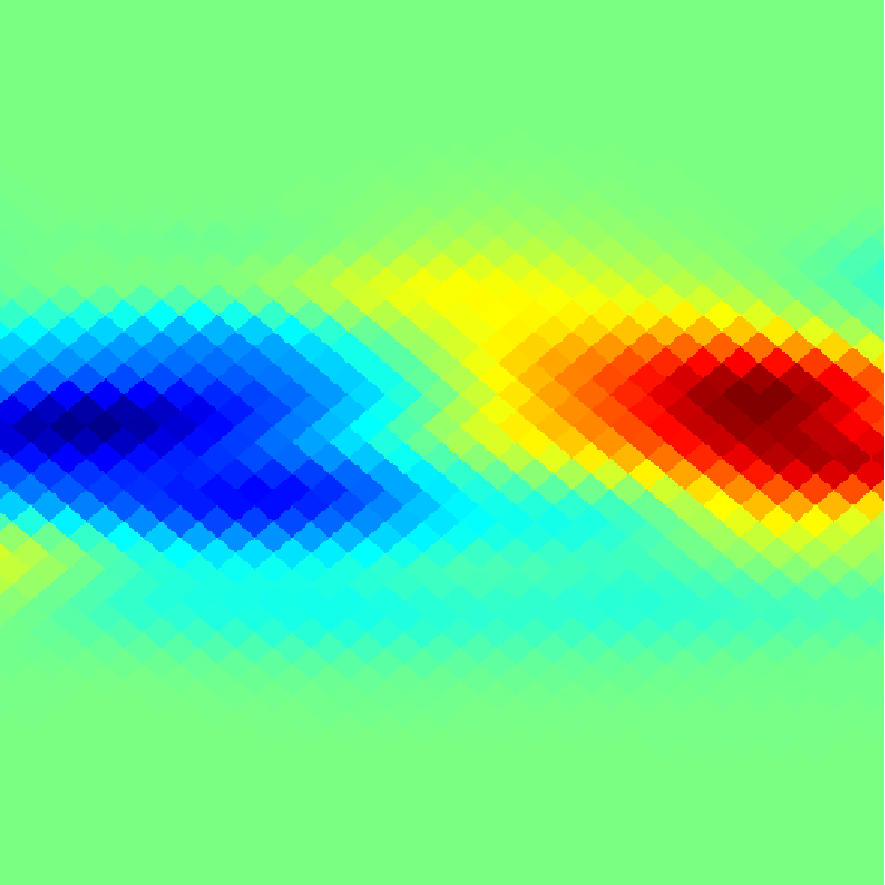}{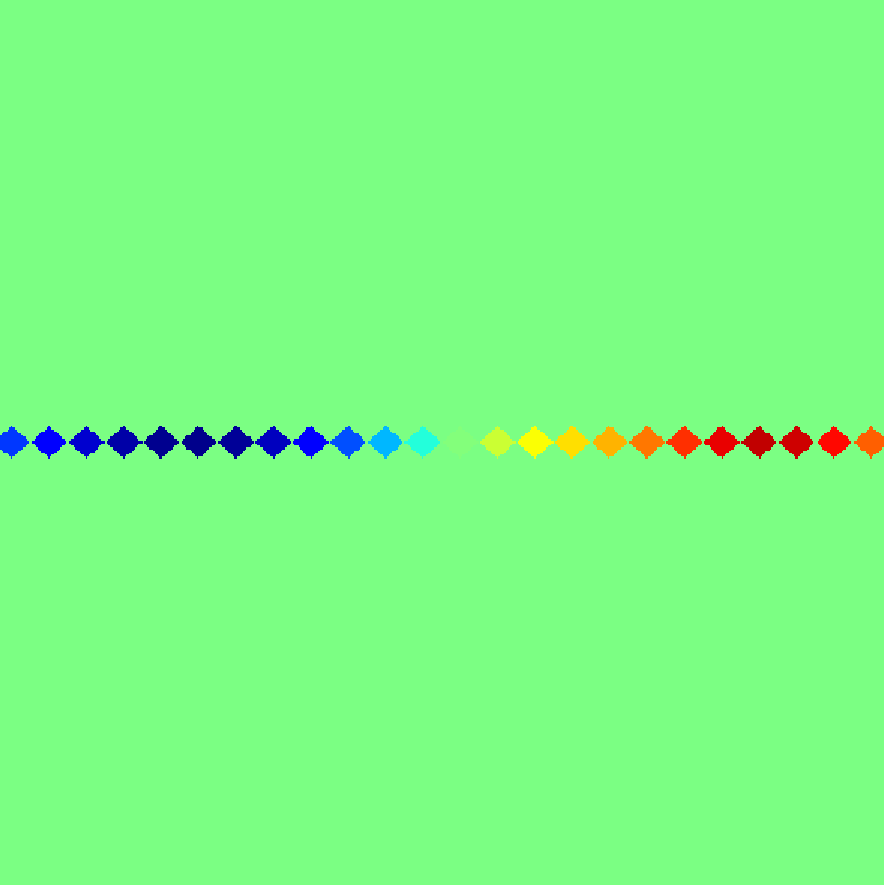}{fig:method}%
{Illustration of the convolution algorithm. To smooth a specific ring
  of the input map (upper left), we first interpolate the kernel on
  the grid, centered on this particular ring (upper right). After a
  ringwise Fourier transformation of both map and kernel, we multiply
  the two components (lower left, transformed to pixel space for
  illustrative purposes). Finally, we perform the summation alongside
  the longitudinal direction and write the inverse Fourier transformed
  result to the output (lower right image).}

\subsection{Implementation and benchmarks}

The algorithm was implemented in C++ using the NVIDIA CUDA 3.2
programming
toolkit\footnote{\url{http://developer.nvidia.com/object/cuda_3_2_downloads.html}}
to generate GPU-related code sequences. We also implemented the hybrid
algorithm for the CPU mostly for validation purposes. To obtain a
program that runs efficiently on GPUs, a basic understanding of the
hardware properties is necessary. We therefore discuss some of the
relevant aspects in the following.

GPUs are streaming devices designed for highly parallel high
throughput data processing. The hardware used here, a NVIDIA GeForce
GTX 480, is a consumer GPU featuring 15 multiprocessors with a total
of 480 shader processors. It is equipped with 1.5~GB of memory and has
a nominal peak performance of 1.3 TFLOP/s in single precision. The
value for double precision arithmetic would be half as large but has
been intentionally degraded by an additional factor of four by the
vendor. For comparison, we note that the performance of the quad-core
CPU used for our benchmark tests is about 45 GFLOP/s.

The latency of main memory accesses with a theoretical bandwidth of
177~GB/s is large: we typically expect a delay of several hundred
clock cycles from the fetch command to the point where the requested
datum is actually available. In the latest generation of NVIDIA GPUs,
a L1 cache of up to 48~KB, dedicated to a specific multiprocessor, and
a global L2 cache of 768~KB may reduce the memory latency. Besides
main memory, the GPU offers up to 48~KB of low latency shared and
64~KB of cached constant memory. In addition, at most 63 registers
with virtually instantaneous access and the highest data throughput
rate are available per thread.

In general, there are two means of hiding memory latencies:
thread-level parallelism and instruction-level parallelism. On GPUs,
threads are very lightweight and the switching between them is
fast. Common practice is therefore to divide the work load over
considerably more threads than physically available computing
cores. The second strategy is less obvious. It attempts to calculate
several unconditional outputs within the same thread. If a thread
encounters a cache miss while computing result A, it can partially
hide the latency by continuing to work on the independent result
B. Our tests show that reducing the total number of active threads by
exploiting instruction-level parallelism enhances code performance by
several tens of percent.

Computations on the GPU are triggered by launching \textit{kernels}
(not to be confused with ``convolution kernels''). As a parameter to
such a function call, we have to specify how the work load should be
processed. More precisely, we define a grid consisting of 1D or 2D
blocks that are consecutively assigned to a GPU multiprocessor. Each
block contains information about the number of threads to be executed
on the device.

Our hybrid algorithm can be implemented in a form that is particularly
well suited to calculations on GPUs. After transferring the regridded
input map to device memory, we use the CUFFT library of toolkit
version 3.1 to compute a ringwise out-of-place real-to-complex FFT on
the input map. In contrast to claims in the vendor's release notes
accompanying the latest version 3.2, we found the predecessor of the
algorithm to be more efficient in terms of both computational
performance and memory consumption, though it proves less accurate for
certain transformation lengths.

To interpolate the convolution kernel on the grid, we launch a GPU
kernel comprising a two-dimensional grid of $\nside / 512 \times
n_{\mathrm{support}}$ blocks with 128 threads. Each thread computes
the output for 8 different pixels within a ring, with a stride given
by the block width. For an efficient interpolation, we stored the
longitudes of all rings and the interpolated convolution kernel in
constant memory. This code section is compute bound and can be
accelerated by taking advantage of the grid symmetries. We calculate
the kernel only on the first $2 \, \nside + 1$ pixels of a ring and
only for the northern hemisphere.

After the function call has been completed, we execute a
real-to-complex batch FFT on the kernel map.

The reduction along the $\theta$-direction is computed partially using
thread-save atomic add operations, available on GPUs of compute
capability 2.0 and higher. We launch a GPU kernel with a
two-dimensional grid of $\lceil (2 \, \nside + 1) / 64 \rceil \times
\lceil n_{\mathrm{support}} / 32 \rceil$ blocks with $64 \times 2$
threads. Each thread accumulates the product of input map and kernel
on up to 32 rings in a local variable and adds the result via an
atomic operation to the global output map. A block performs this
calculation on the northern and southern hemisphere
simultaneously. Although a sophisticated design pattern for the common
reduction problem exists, we found this approach to be more efficient
because we have to deal with non-contiguous memory accesses. This
function is memory bound, but we reach up to 145~GB/s sustained memory
throughput, above 80~\% of the theoretically achievable peak
performance.

We finally compute an inverse complex-to-real FFT before we transfer
the data back to host memory.

In contrast to the exact solution, we find the error in the first and
last ring to be unexpectedly enhanced. This is probably the result of
amplified numerical errors. Although more of a cosmetic correction
than one motivated by accuracy considerations, we recalculate the
values of these eight pixels via a direct sum.

With an implementation as described above, a significant speedup
compared to a FSHT-based convolution can be realized for compact
kernels. We show the results of our benchmark tests in
\fig{fig:scaling}, where we compare the runtime of our algorithm for
different map resolution parameters and kernel sizes to that of the
HEALPix FSHT, and to the optimized \libpsht\ library on a single CPU
core. We note that the GPU timings include the time for transferring
the input map from host memory to GPU memory and the output map from
the GPU back to the host.

We observe a dependence of the performance gain on the resolution
parameter $\nside$. For comparatively low-resolution maps and
small-to-moderate kernel sizes, we find the poorer numerical
efficiency to be the result of too small a workload to enable the
efficient use of the available GPU hardware resources. In addition, we
explicitly optimized our code for fast convolutions at $\nside =
2048$. For a kernel support of $1^{\circ}$, the runtime is completely
dominated by the computational costs of the FFTs. However, with an
increasing kernel size, the scaling behavior changes as both the
kernel evaluation and the multiplication of the Fourier coefficients
of kernel and map become more and more expensive. Accordingly, the
higher efficiency that we achieve flattens out towards higher kernel
diameters. We specify the fractional computational costs of the
different code sections for the convolution at $\nside = 2048$ in
\tab{tab:cputime}.  Since the three major parts of the algorithm take
up comparable amounts of GPU time, further implementation
optimizations for the GPU are unlikely to result in performance gains
significantly larger than the 10~\% level.

In the case of the narrow convolution kernels often encountered during
CMB map beam-convolution processes, performance improvements of up to
two orders of magnitudes can be achieved. For example, smoothing a
HEALPix map with resolution $\nside = 2048, \lmax = 4096$ using a
Gaussian kernel of $4.7\arcmin$ full width at half maximum (FWHM), a
realistic value for the Planck 217~GHz channel
\citep{2011arXiv1101.2048P} takes about $t_{\mathrm{ARKCoS}} = 2.2~s$
on the GPU, whereas the FSHT-based approach requires
$t_{\mathrm{HEALPix}} = 460~s$ and $t_{\mathrm{libpsht}} = 160~s$ on
one CPU core. Although the intrinsically parallel structure of the
algorithm can be most beneficially exploited when run on GPUs, a
CPU-based implementation may also be appropriate for very compact
kernels. For the setup discussed above, the convolution takes about
$t_{\mathrm{ARKCoS}}^{\mathrm{CPU}} = 20~s$ on one single CPU core,
which is still considerably faster than the FSHT-based code.

The cost of realizing these performance gains is to add a GPU at about
half of the cost of the quad-core host system. To compare performance
per hardware dollar, the GPU timings should be compared to half the
CPU timings.

\onefig{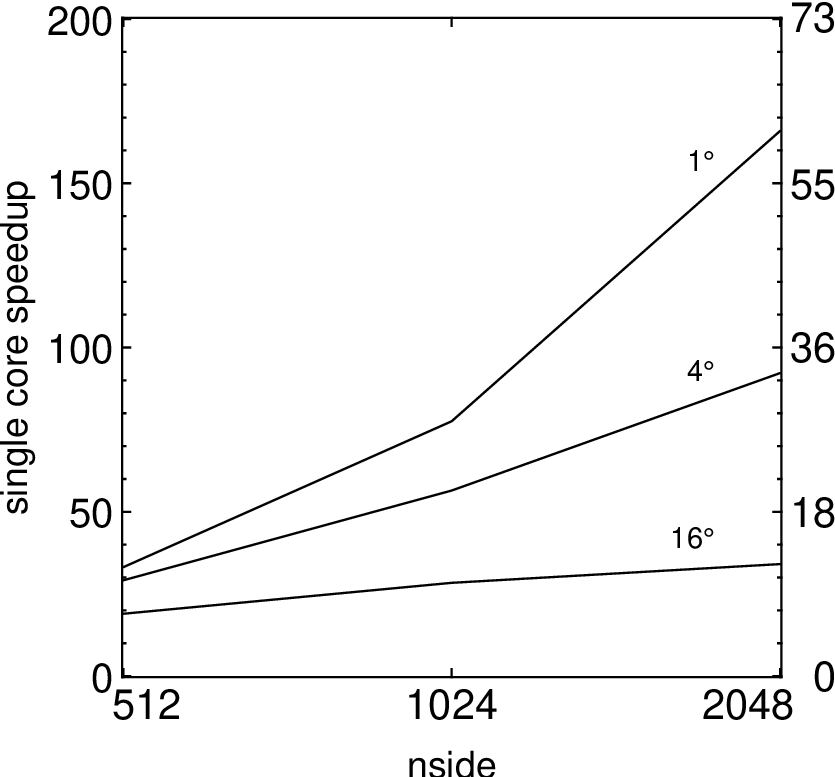}{fig:scaling}%
{Performance gain of the GPU-based convolution code when compared to
  the HEALPix FSHT-based implementation (left axis) and the \libpsht\
  library (right axis) running on a single CPU core for different map
  resolution parameters and kernel support.}

\begin{table}
  \centering
  \caption{Breakdown of the total runtime into the contributions of
    the three most important code sections for the convolution of a
    map at $\nside = 2048$ with kernels of various sizes.}
  \label{tab:cputime}
  \begin{tabular}{c r r r}
    \hline
    \hline
    Kernel support & $1^{\circ} \ $ & $4^{\circ} \ $ & $16^{\circ} \ $\\
    \hline
    FFTs & 57~\% & 47~\% & 39~\% \\
    Kernel evaluation & 19~\% & 31~\% & 39~\% \\
    Ring reduction & 10~\% & 16~\% & 21~\% \\
    Others & 14~\% & 6~\% & 1~\% \\
    \hline
    \hline
  \end{tabular}
\end{table}

\subsection{Accuracy tests}

Our accuracy goal was to achieve a fractional root mean square (RMS)
accuracy of \order{10^{-4}} or lower, which would be sufficient for
most CMB applications. We assessed the accuracy of the newly developed
algorithm on the basis of both the pixel space representation of the
convolved maps, and their power spectra.

For the first test, we computed difference maps of the output
generated by ARKCoS and HEALPix. Using a Gaussian kernel with
$1^{\circ}$~FWHM, as plotted in \fig{fig:kernel}, we show the result
of the comparison in \fig{fig:accuracy}. We note that we normalized
the difference using the RMS of the reference map to obtain a relative
percentage error. We find the small remaining residual around the
polar caps to be dominated by outliers produced by the FFT library for
specific transformation lengths\footnote{Tests with the identical code
  linked to the more recent CUFFT library version 3.2 made these
  outliers disappear, but performance suffered at the 15~\% level.},
whereas inaccuracies in the kernel evaluation prevail in the
equatorial region. Averaged over the entire map, ARKCoS reproduces the
results from the HEALPix package for different kernels with a
fractional RMS error of at most \order{10^{-4}}, which decreases
rapidly for kernel sizes $\gtrsim 0.5^{\circ}$~FWHM.

As a second test, we compared the power spectrum of the convolved map
with the theoretical expectation. With a FWHM of $6 \arcmin$, we chose
a very narrow Gaussian kernel close to the grid resolution at $\nside
= 2048$ that is no longer band-limited at $\lmax = 4096$. The
reference power spectrum used in this test was calculated exactly from
the spherical harmonic representations of input map and kernel. For
one realization, the result is shown in the left-hand panel of
\fig{fig:powerspectrum}, and interpreted in terms of the cross-power
spectrum between the map and the induced error.

In addition, we show the power spectrum of the difference map, where
we again compare the output of our algorithm to the exact
solution. Here, the reference map was derived via \eq{eq:lspacedef}
from the spherical harmonic coefficients of input map and kernel. For
one realization, we show the result in the right-hand panel of
\fig{fig:powerspectrum}, it represents the auto power spectrum of the
error. In both cases, we find the error in the power spectra to be
subdominant over the full dynamical range of about 14
magnitudes\footnote{Note that in \fig{fig:powerspectrum}, we show the
  power spectra multiplied by $\ell \, (\ell + 1) / 2 \pi$. This
  factor has to be taken into account when the dynamical range of the
  simulation is to be assessed.}, showing that the algorithm does not
introduce a significant level of artificial mode coupling.

We conclude with the remark that highly compact kernels of
scale-lengths smaller than $\approx 10 \arcmin$~FWHM, a regime of
particular relevance to beam convolution at the resolution of the
Planck high frequency instrument, will suffer from truncation errors
if a band-limit of $\lmax = 4096$ is imposed. As a result, the
back-transformed pixel space representation starts to show ringing
artifacts. In contrast to FSHT-based algorithms, it is possible to
suppress this effect in our convolution scheme. The first modification
of the algorithm concerns the treatment of super-Nyquist modes in the
polar caps. These modes are available to us because we supersample the
kernel in direct space on $4 \, \nside$ points on all rings. After
performing the forward Fourier transform of the input map, we now
duplicate the coefficients to obtain a fully populated rectangular
grid with $2 \, \nside + 1$ elements on all rings. Likewise, we add
the super-Nyquist modes to the sub-Nyquist modes prior to calculate
the inverse Fourier transform of the output map. This (optional) step
in the algorithm adds a factor of less than two to the computational
time. In the equatorial region, the error can be removed completely if
we slightly alter the algorithm on every unshifted ring. Here, we
start the convolution using the unmodified Fourier transform of the
input map, that is, we do not apply \eq{equ:fftrot}. We instead take
into account the offset on every other ring during the kernel
evaluation, i.e., we substitute $\cos(\phi)$ with $\cos(\phi -
\phi_{\mathrm{0}})$ in \eq{eq:kernel_angle}. In \fig{fig:ringing}, we
compare the output of our modified algorithm to that of a FSHT-based
scheme for the convolution of several point sources with a Gaussian
beam of width $4.7\arcmin$~FWHM at $\nside = 2048$. The conventional
approach suffers from spurious ringing effects that extend well beyond
the formal support of the kernel. Using ARKCoS, the artifacts are
completely absent in the equatorial region, and suppressed and
confined to the latitudinal direction in the polar caps. We note that
on the more regular ECP-grid, the ringing pattern would vanish exactly
on the entire sphere without the need to modify the algorithm.

\twofig{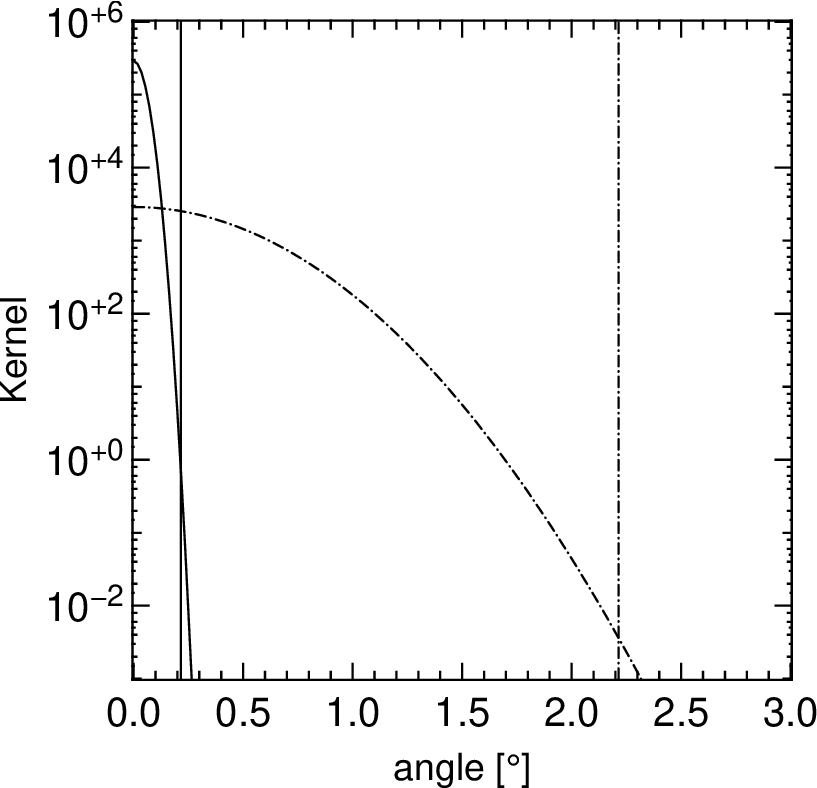}{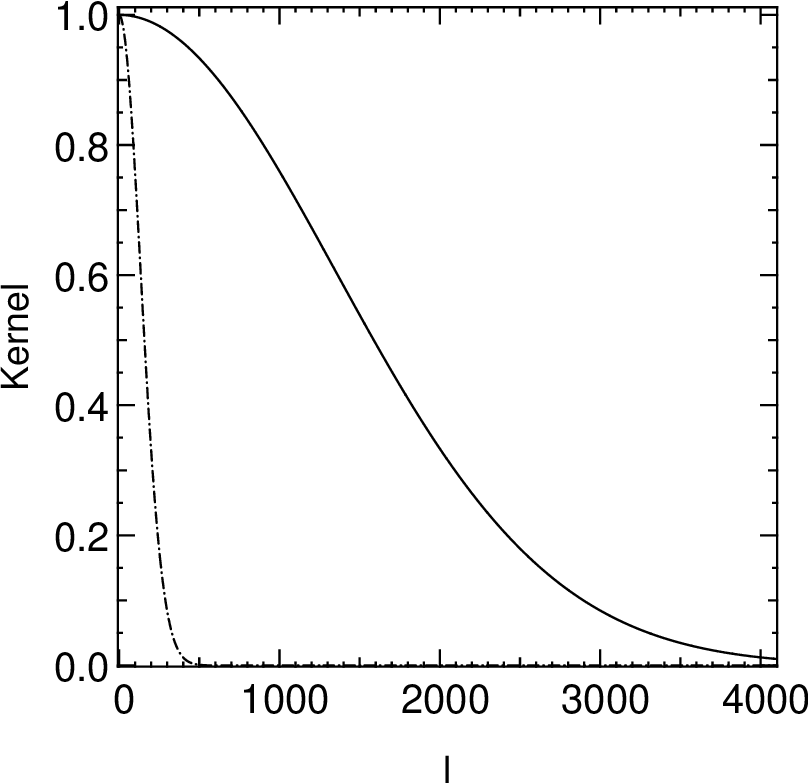}{fig:kernel}%
{Profile of the kernels used for our accuracy tests. Left panel: we
  show the pixel space representation of a Gaussian kernel with
  $6\arcmin$ (solid line) and $1^{\circ}$~FWHM (dashed line). The
  vertical lines indicate the angle beyond which the kernel is
  truncated. Right panel: Legendre expansion of the kernels (solid
  line: $6\arcmin$~FWHM, dashed line: $1^{\circ}$~FWHM). Kernels that
  are compact in pixel space cover a wide range of modes in spherical
  harmonic space and vice versa.}

\twofig{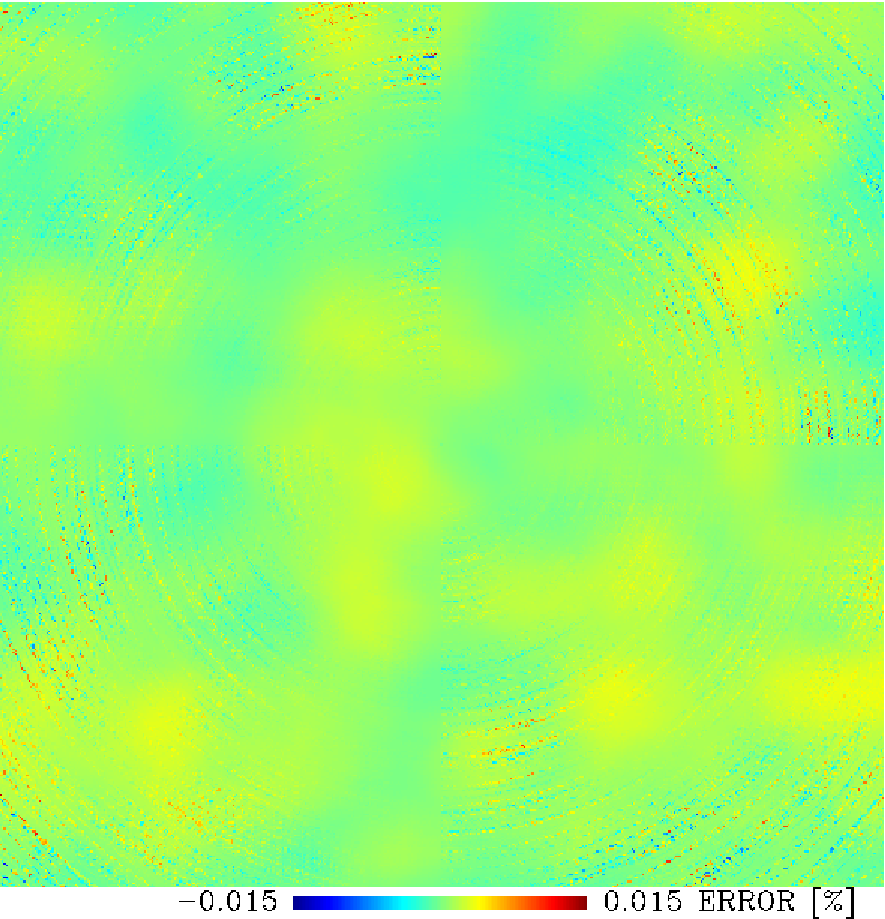}{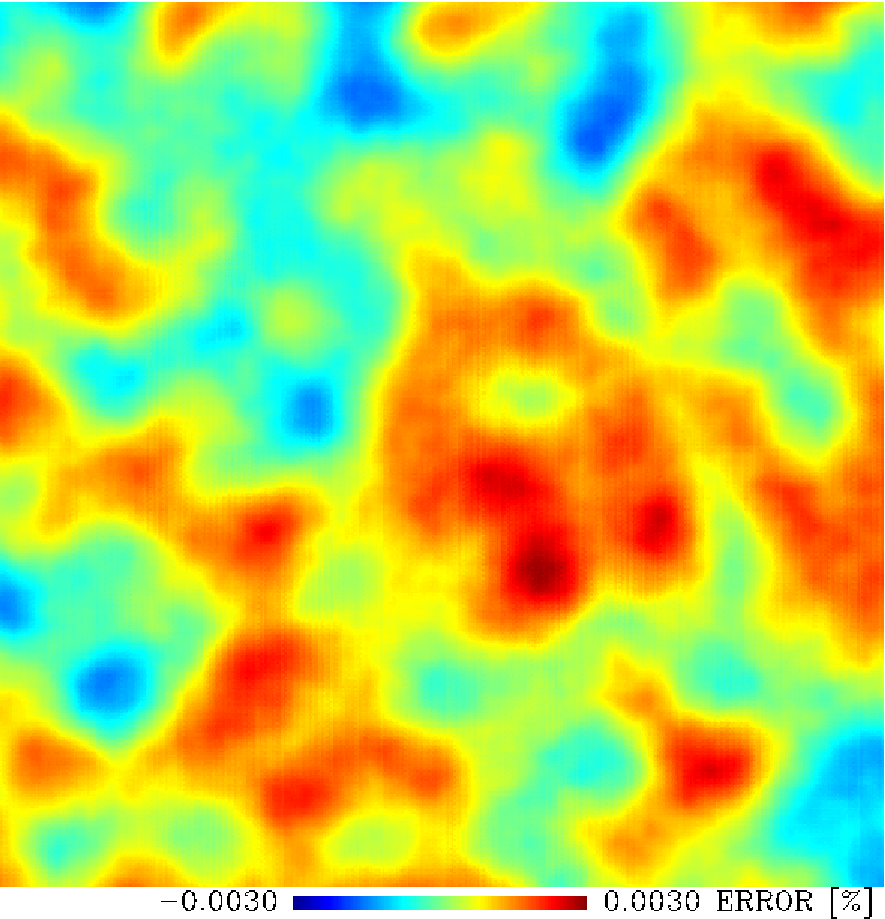}{fig:accuracy}%
{Result of the accuracy test. The difference between a map smoothed in
  spherical harmonic space and a map smoothed with our hybrid method
  is at most $1.5\times 10^{-4}$ in a small number of outliers around
  the north pole (left panel). These are generated by numerical errors
  in the CUFFT library v. 3.1 for specific HEALPix ring lengths and
  are absent when using the slightly slower CUFFT v. 3.2. The larger
  scale \order{10^{-5}} error both at the pole and in the equatorial
  region (right panel) is caused by small inaccuracies in the kernel
  evaluation. In this test, a Gaussian kernel with $1^{\circ}$~FWHM
  was used. Each patch is $10^{\circ}$ on the side.}

\twofig{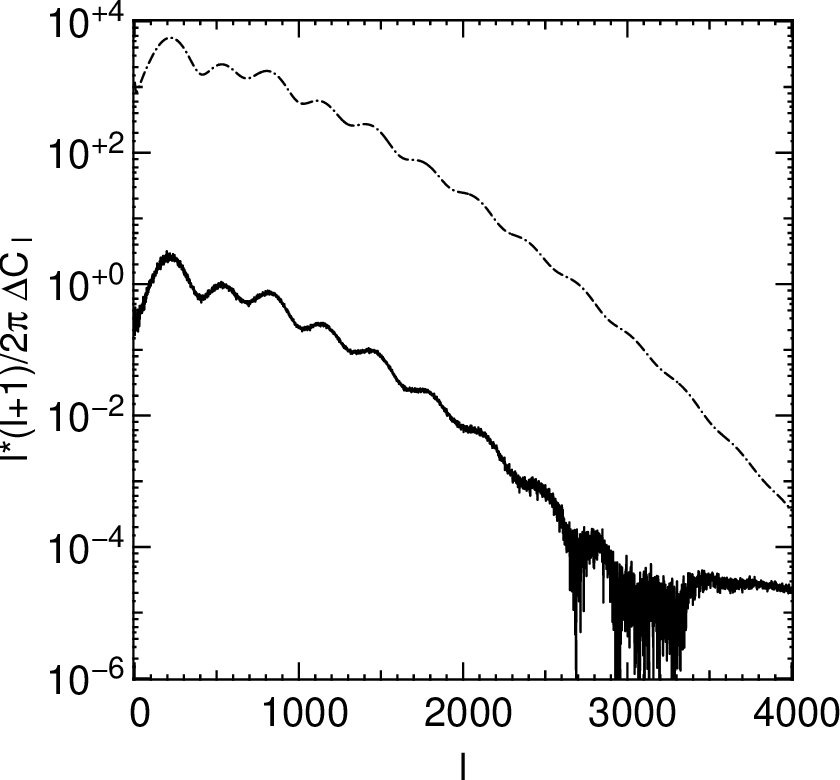}{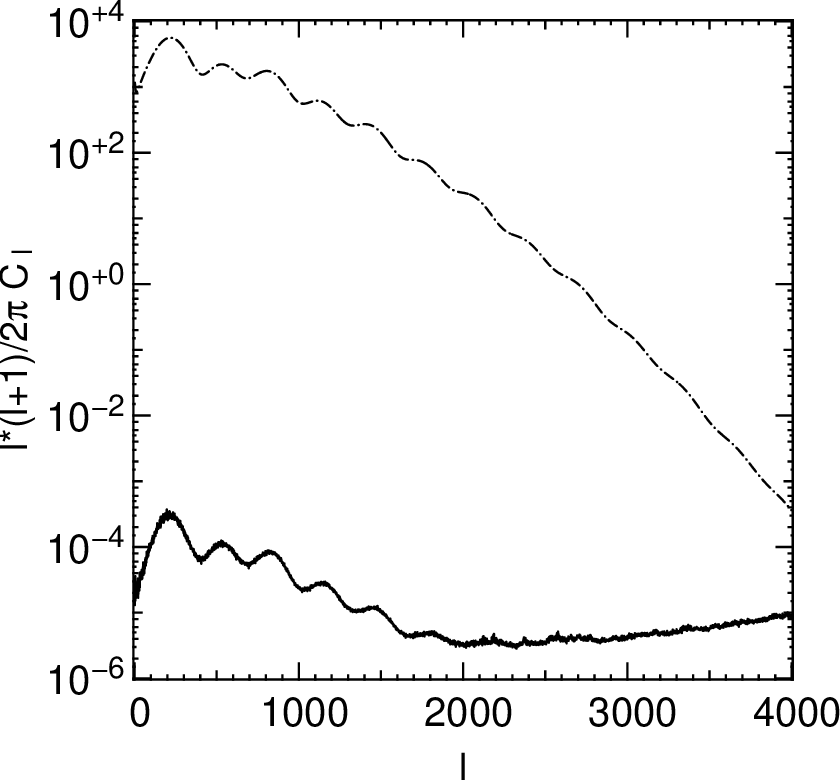}{fig:powerspectrum}%
{Power spectrum accuracy. Left panel: for one particular realization,
  we plot the difference between the power spectrum of a map convolved
  with a narrow Gaussian, FWHM=$6 \arcmin$ and the power spectrum of
  the exact convolution. Right panel: we show the power spectrum of
  the difference map, computed from the convolution of a map using
  ARKCoS and the exact solution. For comparison, we also show the
  expected power spectrum of the exact convolution in both panels
  (dashed lines). Note that since we are comparing to the exact
  convolved power spectrum this error measure includes the HEALPix
  quadrature error.}

\twofig{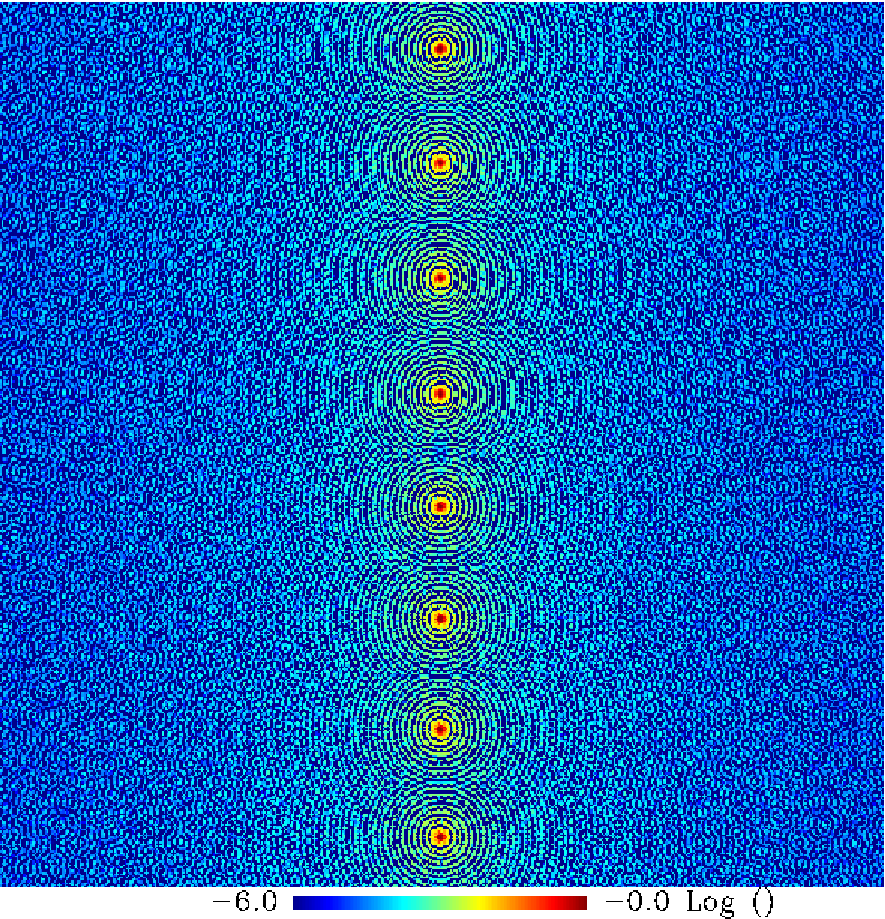}{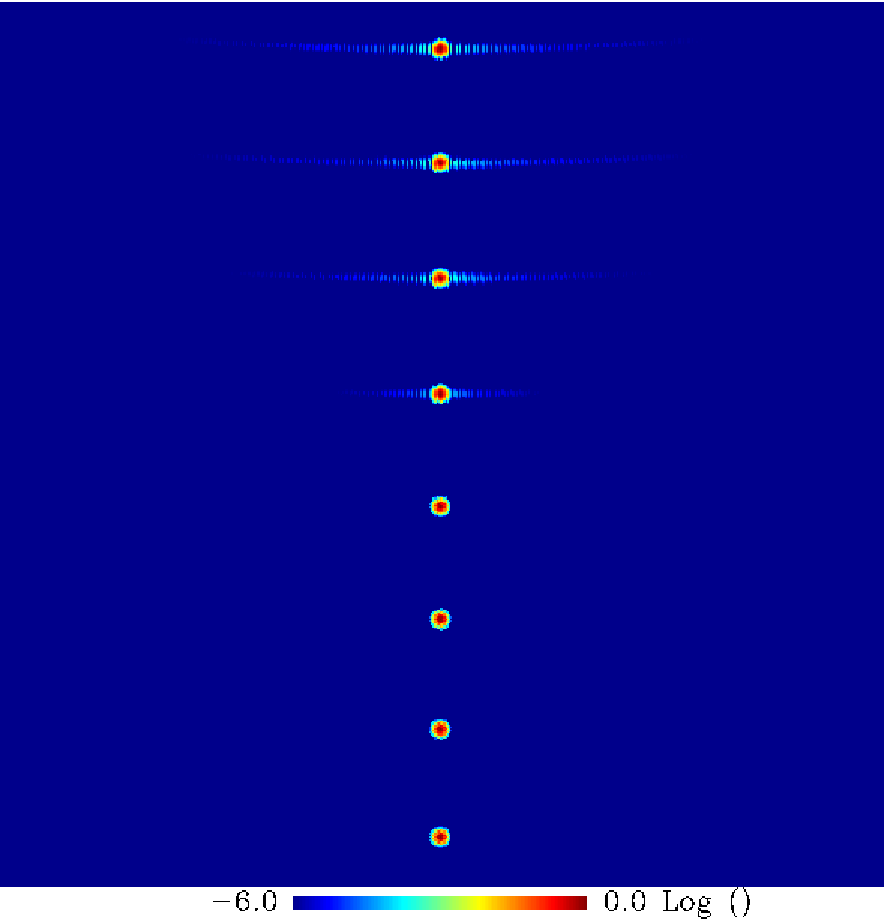}{fig:ringing}%
{Reduced ringing-artifacts in our enhanced hybrid algorithm. Left
  panel: the convolution of point sources with a Gaussian kernel of
  $4.7\arcmin$~FWHM using a FSHT-based algorithm at $\nside = 2048$
  causes extended residuals. Right panel: the result obtained with
  ARKCoS only shows a suppressed ringing pattern in the longitudinal
  direction in the polar caps (upper four point sources). In the
  equatorial region, the artifacts cancel out exactly (lower four
  point sources). Each patch is $13^{\circ}$ on the side, the
  logarithmic color scale counts representing factors of 10 from the
  maximum. For ARKCoS, the ringing patterns are too small to be
  visible on a linear scale.}

\section{Discussion and conclusion}
\label{sec:conclusions}

We have presented an implementation of a GPU-accelerated hybrid
algorithm for radial kernel convolution on the sphere. It performs the
convolution along isolatitude rings in Fourier space and integrates in
longitudinal direction in pixel space. We call this algorithm ARKCoS.
As the computational costs scale linearly with the kernel support, the
method is most beneficial for convolution with compact
kernels. Typical applications include CMB beam smoothing, symmetric
wavelet analyses, and point-source filtering operations.

For a convolution with compact kernels, we find that our
implementation realizes real performance gains of up to 5000~\%,
depending on the problem size, for a 50~\% increase in system cost
relative to the most widely used FSHT implementation in the HEALPix
library running in parallel on a quad-core CPU. When compared to the
more finely tuned \libpsht\ FSHT library, again running on four cores,
we still find significant performance gains, up to 1800~\%.

We assessed the numerical accuracy of the algorithm by comparing the
convolved output map to the result generated using HEALPix. The
outcome typically agrees with the FSHT-based convolution to 1 part in
$10^{4}$. Comparing the power spectrum of the output map to the exact
solution for a narrow convolution kernel, we find a relative error of
smaller than $10^{-3}$. For kernels that are not band-limited, the
convolution with a FSHT-based scheme induces ringing artifacts. Using
instead a slightly modified implementation of ARKCoS, however, we have
demonstrated that a huge reduction in the spurious contribution is
possible.

The massively parallel hybrid approach we have presented here is
particularly advantageous for convolutions with compact kernels (with
support less than $\sim 15^{\circ}$) at high resolution ($\nside =
512$ or higher). The GPU we used for our tests has 1.5 GB of RAM. This
is too small to store simultaneously the input and output HEALPix maps
at resolution $\nside = 4096$ or higher. Possible solutions for future
work involve calculating the contribution to the output map from
subsets of rings in the input map, either sequentially or in parallel
if more than one GPU is available in the same system.

This work deals with radial kernel convolution. We note in closing
that there is considerable interest in the algorithmically more
difficult problem of asymmetric kernel convolution
(\citealt{2001PhRvD..63l3002W, 2005AGUFMIN23A1201W}; ASYMFAST,
\citealt{2004PhRvD..69l3008T}; FICSBell, \citealt{ficsbell}; FEBECoP,
\citealt{2011ApJS..193....5M}) either to model the physical optics of
CMB experiments more faithfully \citep{2011arXiv1101.2038M,
  2011arXiv1101.2048P} or to detect signals that have locally
anisotropic signatures. Having found in this work that our hybrid
algorithm vastly accelerates radial kernel convolution, it is easy to
imagine generalizations that accelerate asymmetric kernel convolution
in a similar way. The ASYMFAST approach \citep{2004PhRvD..69l3008T}
reduces the problem of asymmetric beam convolution to \order{10}
symmetric convolutions. Coupled to our GPU accelerated approach, the
convolution with even complex asymmetric kernels and compact support
takes less time than the convolution with a single symmetric kernel on
a CPU system using fast spherical harmonic transforms.

\begin{acknowledgements}
  We thank the anonymous referee for the comments which helped to
  improve the presentation of our results. We are grateful to
  S. Kruger for some early exploratory GPU work. We thank F. Bouchet,
  C. Lawrence, and S. Gratton for their encouragement at various
  stages during the project, and E. Hivon for comments on the draft of
  this paper. BDW was supported by NASA/JPL subcontract 1413479, and
  NSF grants AST 07-08849, AST 09-08693 ARRA and AST 09-08902 during
  this work. This research was supported in part by the National
  Science Foundation through TeraGrid resources provided by NCSA under
  grant number TG-AST100029. Some of the results in this paper have
  been derived using the HEALPix \citep{2005ApJ...622..759G}, the
  \libpsht\ \citep{2011A&A...526A.108R}, and the FFTW \citep{FFTW05}
  packages.
\end{acknowledgements}

\bibliographystyle{aa}
\bibliography{literature}

\begin{thebibliography}{28}
\expandafter\ifx\csname natexlab\endcsname\relax\def\natexlab#1{#1}\fi

\bibitem[{{Baumann} {et~al.}(2009){Baumann}, {Jackson}, {Adshead}, {Amblard},
  {Ashoorioon}, {Bartolo}, {Bean}, {Beltr\'{a}n}, {de Bernardis}, {Bird},
  {Chen}, {Chung}, {Colombo}, {Cooray}, {Creminelli}, {Dodelson}, {Dunkley},
  {Dvorkin}, {Easther}, {Finelli}, {Flauger}, {Hertzberg}, {Jones-Smith},
  {Kachru}, {Kadota}, {Khoury}, {Kinney}, {Komatsu}, {Krauss}, {Lesgourgues},
  {Liddle}, {Liguori}, {Lim}, {Linde}, {Matarrese}, {Mathur}, {McAllister},
  {Melchiorri}, {Nicolis}, {Pagano}, {Peiris}, {Peloso}, {Pogosian},
  {Pierpaoli}, {Riotto}, {Seljak}, {Senatore}, {Shandera}, {Silverstein},
  {Smith}, {Vaudrevange}, {Verde}, {Wandelt}, {Wands}, {Watson}, {Wyman},
  {Yadav}, {Valkenburg}, \& {Zaldarriaga}}]{baumann:10}
{Baumann}, D., {Jackson}, M.~G., {Adshead}, P., {et~al.} 2009, AIP Conference
  Proceedings, 1141, 10

\bibitem[{{Cay{\'o}n} {et~al.}(2000){Cay{\'o}n}, {Sanz}, {Barreiro},
  {Mart{\'{\i}}nez-Gonz{\'a}lez}, {Vielva}, {Toffolatti}, {Silk}, {Diego}, \&
  {Arg{\"u}eso}}]{2000MNRAS.315..757C}
{Cay{\'o}n}, L., {Sanz}, J.~L., {Barreiro}, R.~B., {et~al.} 2000, \mnras, 315,
  757

\bibitem[{{Doroshkev} {et~al.}(2005){Doroshkev}, {Naselsky}, {Verkhodanov},
  {Novikov}, {Turchaninov}, {Novikov}, {Christensen}, \&
  {Chiang}}]{2005IJMPD..14..275D}
{Doroshkev}, A.~G., {Naselsky}, P.~D., {Verkhodanov}, O.~V., {et~al.} 2005,
  International Journal of Modern Physics D, 14, 275

\bibitem[{{Driscoll} \& {Healy}(1994)}]{184069.184073D}
{Driscoll}, J.~R. \& {Healy}, D.~M. 1994, Adv. Appl. Math., 15, 202

\bibitem[{{Frigo} \& {Johnson}(2005)}]{FFTW05}
{Frigo}, M. \& {Johnson}, S.~G. 2005, Proceedings of the IEEE, 93, 216

\bibitem[{{Gonz{\'a}lez-Nuevo} {et~al.}(2006){Gonz{\'a}lez-Nuevo},
  {Arg{\"u}eso}, {L{\'o}pez-Caniego}, {Toffolatti}, {Sanz}, {Vielva}, \&
  {Herranz}}]{2006MNRAS.369.1603G}
{Gonz{\'a}lez-Nuevo}, J., {Arg{\"u}eso}, F., {L{\'o}pez-Caniego}, M., {et~al.}
  2006, \mnras, 369, 1603

\bibitem[{{G{\'o}rski} {et~al.}(2005){G{\'o}rski}, {Hivon}, {Banday},
  {Wandelt}, {Hansen}, {Reinecke}, \& {Bartelmann}}]{2005ApJ...622..759G}
{G{\'o}rski}, K.~M., {Hivon}, E., {Banday}, A.~J., {et~al.} 2005, \apj, 622,
  759

\bibitem[{{Hivon} \& {Ponthieu}(2011)}]{ficsbell}
{Hivon}, E. \& {Ponthieu}, N. 2011, in preparation

\bibitem[{{Hobson} {et~al.}(1999){Hobson}, {Jones}, \&
  {Lasenby}}]{1999MNRAS.309..125H}
{Hobson}, M.~P., {Jones}, A.~W., \& {Lasenby}, A.~N. 1999, \mnras, 309, 125

\bibitem[{{Hupca} {et~al.}(2010){Hupca}, {Falcou}, {Grigori}, \&
  {Stompor}}]{2010arXiv1010.1260H}
{Hupca}, I.~O., {Falcou}, J., {Grigori}, L., \& {Stompor}, R. 2010, ArXiv
  e-prints

\bibitem[{{Kosowsky}(2003)}]{2003NewAR..47..939K}
{Kosowsky}, A. 2003, \nar, 47, 939

\bibitem[{{Mart{\'{\i}}nez-Gonz{\'a}lez}
  {et~al.}(2002){Mart{\'{\i}}nez-Gonz{\'a}lez}, {Gallegos}, {Arg{\"u}eso},
  {Cay{\'o}n}, \& {Sanz}}]{2002MNRAS.336...22M}
{Mart{\'{\i}}nez-Gonz{\'a}lez}, E., {Gallegos}, J.~E., {Arg{\"u}eso}, F.,
  {Cay{\'o}n}, L., \& {Sanz}, J.~L. 2002, \mnras, 336, 22

\bibitem[{{Mennella} {et~al.}(2011){Mennella}, {Bersanelli}, {Butler}, {Curto},
  {Cuttaia}, {Davis}, {Dick}, {Frailis}, {Galeotta}, {Gregorio},
  {Kurki-Suonio}, {Lawrence}, {Leach}, {Leahy}, {Lowe}, {Maino}, {Mandolesi},
  {Maris}, {Mart$\backslash$'$\backslash$inez-Gonz{\'a}lez}, {Meinhold},
  {Morgante}, {Pearson}, {Perrotta}, {Polenta}, {Poutanen}, {Sandri},
  {Seiffert}, {Suur-Uski}, {Tavagnacco}, {Terenzi}, {Tomasi}, {Valiviita},
  {Villa}, {Watson}, {Wilkinson}, {Zacchei}, {Zonca}, {Aja}, {Artal},
  {Baccigalupi}, {Banday}, {Barreiro}, {Bartlett}, {Bartolo}, {Battaglia},
  {Bennett}, {Bonaldi}, {Bonavera}, {Borrill}, {Bouchet}, {Burigana},
  {Cabella}, {Cappellini}, {Chen}, {Colombo}, {Cruz}, {Danese}, {D'Arcangelo},
  {Davies}, {de Gasperis}, {de Rosa}, {de Zotti}, {Dickinson}, {Diego},
  {Donzelli}, {Efstathiou}, {En$\backslash$sslin}, {Eriksen}, {Falvella},
  {Finelli}, {Foley}, {Franceschet}, {Franceschi}, {Gaier},
  {G{\'e}nova-Santos}, {George}, {G{\'o}mez}, {Gonz{\'a}lez-Nuevo},
  {G{\'o}rski}, {Gruppuso}, {Hansen}, {Herranz}, {Herreros}, {Hoyland},
  {Hughes}, {Jewell}, {Jukkala}, {Juvela}, {Kangaslahti}, {Keih{\"a}nen},
  {Keskitalo}, {Kilpia}, {Kisner}, {Knoche}, {Knox}, {Laaninen},
  {L{\"a}hteenm{\"a}ki}, {Lamarre}, {Leonardi}, {Le{\'o}n-Tavares},
  {Leutenegger}, {Lilje}, {L{\'o}pez-Caniego}, {Lubin}, {Malaspina},
  {Marinucci}, {Massardi}, {Matarrese}, {Matthai}, {Melchiorri}, {Mendes},
  {Miccolis}, {Migliaccio}, {Mitra}, {Moss}, {Natoli}, {Nesti},
  {N$\backslash$orgaard-Nielsen}, {Pagano}, {Paladini}, {Paoletti},
  {Partridge}, {Pasian}, {Pettorino}, {Pietrobon}, {Pospieszalski},
  {Pr{\'e}zeau}, {Prina}, {Procopio}, {Puget}, {Quercellini}, {Rachen},
  {Rebolo}, {Reinecke}, {Ricciardi}, {Robbers}, {Rocha}, {Roddis},
  {Rubi$\backslash$no-Mart{\'{\i}}n}, {Savelainen}, {Scott}, {Silvestri},
  {Simonetto}, {Sjoman}, {Smoot}, {Sozzi}, {Stringhetti}, {Tauber}, {Tofani},
  {Tuovinen}, {T{\"u}rler}, {Umana}, {Valenziano}, {Varis}, {Vielva},
  {Vittorio}, {Wade}, {Watson}, {White}, \& {Winder}}]{2011arXiv1101.2038M}
{Mennella}, A., {Bersanelli}, M., {Butler}, R.~C., {et~al.} 2011, ArXiv
  e-prints

\bibitem[{{Mitra} {et~al.}(2011){Mitra}, {Rocha}, {G{\'o}rski}, {Huffenberger},
  {Eriksen}, {Ashdown}, \& {Lawrence}}]{2011ApJS..193....5M}
{Mitra}, S., {Rocha}, G., {G{\'o}rski}, K.~M., {et~al.} 2011, \apjs, 193, 5

\bibitem[{{Muciaccia} {et~al.}(1997){Muciaccia}, {Natoli}, \&
  {Vittorio}}]{1997ApJ...488L..63M}
{Muciaccia}, P.~F., {Natoli}, P., \& {Vittorio}, N. 1997, \apjl, 488, L63+

\bibitem[{{Natoli} {et~al.}(2001){Natoli}, {de Gasperis}, {Gheller}, \&
  {Vittorio}}]{2001A&A...372..346N}
{Natoli}, P., {de Gasperis}, G., {Gheller}, C., \& {Vittorio}, N. 2001, \aap,
  372, 346

\bibitem[{{Planck Collaboration} {et~al.}(2011){Planck Collaboration}, {Ade},
  {Aghanim}, {Arnaud}, {Ashdown}, {Aumont}, {Baccigalupi}, {Baker}, {Balbi},
  {Banday}, \& et~al.}]{2011arXiv1101.2022P}
{Planck Collaboration}, {Ade}, P.~A.~R., {Aghanim}, N., {et~al.} 2011, ArXiv
  e-prints

\bibitem[{{Planck HFI Core Team} {et~al.}(2011){Planck HFI Core Team}, {Ade},
  {Aghanim}, {Ansari}, {Arnaud}, {Ashdown}, {Aumont}, {Banday}, {Bartelmann},
  {Bartlett}, {Battaner}, {Benabed}, {Beno{\^i}t}, {Bernard}, {Bersanelli},
  {Bock}, {Bond}, {Borrill}, {Bouchet}, {Boulanger}, {Bradshaw}, {Bucher},
  {Cardoso}, {Castex}, {Catalano}, {Challinor}, {Chamballu}, {Chary}, {Chen},
  {Chiang}, {Church}, {Clements}, {Colley}, {Colombi}, {Couchot}, {Coulais},
  {Cressiot}, {Crill}, {Crook}, {de Bernardis}, {Delabrouille}, {Delouis},
  {D{\'e}sert}, {Dolag}, {Dole}, {Dor{\'e}}, {Douspis}, {Dunkley},
  {Efstathiou}, {Filliard}, {Forni}, {Fosalba}, {Ganga}, {Giard}, {Girard},
  {Giraud-H{\'e}raud}, {Gispert}, {G{\'o}rski}, {Gratton}, {Griffin}, {Guyot},
  {Haissinski}, {Harrison}, {Helou}, {Henrot-Versill{\'e}},
  {Hern{\'a}ndez-Monteagudo}, {Hildebrandt}, {Hills}, {Hivon}, {Hobson},
  {Holmes}, {Huffenberger}, {Jaffe}, {Jones}, {Kaplan}, {Kneissl}, {Knox},
  {Kunz}, {Lagache}, {Lamarre}, {Lange}, {Lasenby}, {Lavabre}, {Lawrence}, {Le
  Jeune}, {Leroy}, {Lesgourgues}, {Lewis}, {Mac{\'{\i}}as-P{\'e}rez},
  {MacTavish}, {Maffei}, {Mandolesi}, {Mann}, {Marleau}, {Marshall}, {Masi},
  {Matsumura}, {McAuley}, {McGehee}, {Melin}, {Mercier}, {Mitra},
  {Miville-Desch{\^e}nes}, {Moneti}, {Montier}, {Mortlock}, {Murphy}, {Nati},
  {Netterfield}, {N$\backslash$orgaard-Nielsen}, {North}, {Noviello},
  {Novikov}, {Osborne}, {Pajot}, {Patanchon}, {Peacocke}, {Pearson},
  {Perdereau}, {Perotto}, {Piacentini}, {Piat}, {Plaszczynski},
  {Pointecouteau}, {Ponthieu}, {Pr{\'e}zeau}, {Prunet}, {Puget}, {Reach},
  {Remazeilles}, {Renault}, {Riazuelo}, {Ristorcelli}, {Rocha}, {Rosset},
  {Roudier}, {Rowan-Robinson}, {Rusholme}, {Saha}, {Santos}, {Savini},
  {Schaefer}, {Shellard}, {Spencer}, {Starck}, {Stolyarov}, {Stompor},
  {Sudiwala}, {Sunyaev}, {Sutton}, {Sygnet}, {Tauber}, {Thum}, {Torre},
  {Touze}, {Tristram}, {Van Leeuwen}, {Vibert}, {Vibert}, {Wandelt}, {White},
  {Wiesemeyer}, {Woodcraft}, {Yurchenko}, {Yvon}, \&
  {Zacchei}}]{2011arXiv1101.2048P}
{Planck HFI Core Team}, {Ade}, P.~A.~R., {Aghanim}, N., {et~al.} 2011, ArXiv
  e-prints

\bibitem[{{Reinecke}(2011)}]{2011A&A...526A.108R}
{Reinecke}, M. 2011, \aap, 526, A108+

\bibitem[{{Ruhl} {et~al.}(2004){Ruhl}, {Ade}, {Carlstrom}, {Cho}, {Crawford},
  {Dobbs}, {Greer}, {Halverson}, {Holzapfel}, {Lanting}, {Lee}, {Leitch},
  {Leong}, {Lu}, {Lueker}, {Mehl}, {Meyer}, {Mohr}, {Padin}, {Plagge}, {Pryke},
  {Runyan}, {Schwan}, {Sharp}, {Spieler}, {Staniszewski}, \&
  {Stark}}]{2004SPIE.5498...11R}
{Ruhl}, J., {Ade}, P.~A.~R., {Carlstrom}, J.~E., {et~al.} 2004, in Presented at
  the Society of Photo-Optical Instrumentation Engineers (SPIE) Conference,
  Vol. 5498, Society of Photo-Optical Instrumentation Engineers (SPIE)
  Conference Series, ed. {C.~M.~Bradford, P.~A.~R.~Ade, J.~E.~Aguirre,
  J.~J.~Bock, M.~Dragovan, L.~Duband, L.~Earle, J.~Glenn, H.~Matsuhara,
  B.~J.~Naylor, H.~T.~Nguyen, M.~Yun, \& J.~Zmuidzinas}, 11--29

\bibitem[{{Stompor} {et~al.}(2002){Stompor}, {Balbi}, {Borrill}, {Ferreira},
  {Hanany}, {Jaffe}, {Lee}, {Oh}, {Rabii}, {Richards}, {Smoot}, {Winant}, \&
  {Wu}}]{2002PhRvD..65b2003S}
{Stompor}, R., {Balbi}, A., {Borrill}, J.~D., {et~al.} 2002, \prd, 65, 022003

\bibitem[{{Tegmark}(1997)}]{1997ApJ...480L..87T}
{Tegmark}, M. 1997, \apjl, 480, L87+

\bibitem[{{Tegmark} \& {de Oliveira-Costa}(1998)}]{1998ApJ...500L..83T}
{Tegmark}, M. \& {de Oliveira-Costa}, A. 1998, \apjl, 500, L83+

\bibitem[{{Tristram} {et~al.}(2004){Tristram}, {Mac{\'{\i}}as-P{\'e}rez},
  {Renault}, \& {Hamilton}}]{2004PhRvD..69l3008T}
{Tristram}, M., {Mac{\'{\i}}as-P{\'e}rez}, J.~F., {Renault}, C., \& {Hamilton},
  J. 2004, \prd, 69, 123008

\bibitem[{{Vielva} {et~al.}(2004){Vielva}, {Mart{\'{\i}}nez-Gonz{\'a}lez},
  {Barreiro}, {Sanz}, \& {Cay{\'o}n}}]{2004ApJ...609...22V}
{Vielva}, P., {Mart{\'{\i}}nez-Gonz{\'a}lez}, E., {Barreiro}, R.~B., {Sanz},
  J.~L., \& {Cay{\'o}n}, L. 2004, \apj, 609, 22

\bibitem[{{Wandelt} \& {G{\'o}rski}(2001)}]{2001PhRvD..63l3002W}
{Wandelt}, B.~D. \& {G{\'o}rski}, K.~M. 2001, \prd, 63, 123002

\bibitem[{{Wiaux} {et~al.}(2005){Wiaux}, {Jacques}, \&
  {Vandergheynst}}]{2005AGUFMIN23A1201W}
{Wiaux}, Y., {Jacques}, L., \& {Vandergheynst}, P. 2005, AGU Fall Meeting
  Abstracts, A1201+

\bibitem[{{Wiaux} {et~al.}(2007){Wiaux}, {Jacques}, \&
  {Vandergheynst}}]{2007JCoPh.226.2359W}
{Wiaux}, Y., {Jacques}, L., \& {Vandergheynst}, P. 2007, Journal of
  Computational Physics, 226, 2359

\end{thebibliography}

\end{document}